\documentclass[pre,aps,showpacs,superscriptaddress,epsf]{revtex4}
\usepackage{graphicx,color}
\usepackage{amsmath}
\usepackage{amssymb}
\usepackage{bm}

\def\sqbull{\vrule height .9ex width .8ex depth -.1ex} 

\begin{document}


\title{Heat transport measurements in turbulent rotating Rayleigh-B\'{e}nard convection} 
\author{Yuanming Liu $\dag$}
\author{Robert E. Ecke}
\affiliation{Center for Nonlinear Studies}
\affiliation{Condensed Matter and Thermal Physics Group \\ Los Alamos National Laboratory, Los
Alamos, NM 87545}

\begin{abstract}
We present experimental heat transport measurements of turbulent Rayleigh-B\'{e}nard convection with rotation
about a vertical axis. The fluid, water with Prandtl number ($\sigma$) about 6, was confined in
a cell which had a square cross section of 7.3 cm$\times$7.3 cm and a height of 9.4 cm. Heat
transport was measured for Rayleigh numbers $2\times 10^5 <$ Ra $ < 5\times 10^8$ and Taylor numbers $0 <$ Ta $< 5\times 10^{9}$. 
We show the variation of normalized heat transport, the Nusselt number, at fixed dimensional rotation
rate $\Omega_D$, at fixed Ra varying Ta, at fixed Ta varying Ra, and at fixed Rossby number Ro.  The scaling of heat transport in the range $10^7$ to about $10^9$ is roughly 0.29 with a Ro dependent coefficient or equivalently is also well fit by a combination of power laws of the form $a\ Ra^{1/5} + b\ Ra^{1/3}$.  The range of Ra is not sufficient to differentiate
single power law or combined power law scaling.  The overall impact of rotation on heat transport in turbulent convection is assessed.

\end{abstract}
\maketitle


\section{Introduction}

Turbulent thermal convection plays a key role in many of the phenomena associated with
geophysical and astrophysical fluid dynamics \cite{Gerkema08} as well as providing a well-posed
problem for the study of fundamental fluid dynamics \cite{Siggia94}. In several
important examples including oceanic deep convection \cite{Gerkema08} and convection in stars \cite{miesch08} 
and giant planets \cite{Aurnou08},
the effects of rotation are critical in determining the nature of the fluid motion. Rotation also provides an additional parameter for understanding the origin of heat transport scaling in turbulent convection, a topic of
tremendous experimental  activity in recent years \cite{Siggia94,AhlersRMP}.
In comparison, the research efforts applied to rotating turbulent convection have been
rather modest arising from the pioneering theoretical work of Chandrasekhar \cite{Chandra53,Chandra61}. Experimental measurements of heat transport in rotating convection include the
seminal work of Rossby \cite{Rossby69} and later studies that also had qualitative flow visualization \cite{Zhong93}. 
Numerical simulations have also had significant impact \cite{Raasch91,Cabot90,Klinger95,Julien96a,Julien96b}. Here we consider both rotating and non-rotating convection and provide new insights into heat transport scaling of rotating convective turbulence.  A short report of some aspects of this work appeared previously \cite{Liu97}, and further studies of velocity fields in rotating convection motivated by this work were also published \cite{Vorobieff02}.   

Rotating Rayleigh-B\'{e}nard convection can be characterized by three dimensionless parameters:
the Rayleigh number Ra which is a measure of buoyant forcing, the Taylor number Ta which
measures the effect of the rotational Coriolis force, and the Prandtl number $\sigma$ which
determines the dominant nonlinearity in convection. These parameters are defined by: 
\begin{equation}
 {\rm Ra} = {g\alpha d^3 \Delta T\over\nu\kappa}, \hskip 0.2in {\rm Ta}
= \left ({2\Omega_D d^2\over\nu}\right )^2,  \hskip 0.2in \sigma = {\nu\over\kappa}
\label{eq:defs}
\end{equation} 
where $g$ is the acceleration of gravity, $\alpha$ is the thermal expansion
coefficient, $\Delta T$ is the temperature difference across  the fluid layer of height $d$,
$\nu$ is the kinematic viscosity, $\kappa$ is the thermal diffusivity, and $\Omega_D$ is the
physical angular rotation rate. The dimensionless rotation frequency $\Omega = \Omega_D
d^2/\nu$ is sometimes used in place of $\Omega_D$ or Ta. Properties of thermal turbulence
can also be affected by the cell geometry characterized by the ratio of a lateral length to a
vertical length. For our square geometry, we define the cell aspect ratio as $\Gamma\equiv l/d$
where $l$ is the lateral size of the cell.

Although Ra, Ta, $\sigma$ and $\Gamma$ completely define the parameter space of rotating
convection, the behavior of different quantities such as heat transport is complicated when one
parameter is  varied while keeping the others constant. For example, as Ra is increased at fixed
Ta, the relative influence of buoyancy and rotation changes, making an evaluation of the
influence of rotation alone difficult. To ease this problem, it is useful to define \cite{Julien96a} an additional parameter, the convective Rossby number Ro 
\begin{equation}
{\rm Ro} = \sqrt{{{\rm Ra}\over\sigma {\rm Ta}}} 
\label{eq:Rossby}
\end{equation}
which is a measure of a characteristic buoyant velocity to a rotational velocity.  This definition is equivalent to those used previously \cite{Gilman77,Cabot90} and is closely related to other definitions of convective Rossby number \cite{Raasch91,Fernando91,Jones93}.  Roughly speaking, the border between rotation-dominated and buoyancy-dominated flows should be approximated by the condition Ro = 1. 

The quantity of interest here is turbulent heat transport as measured by the Nusselt number Nu, the total heat transported by convection normalized by the heat transported by molecular diffusion alone.  To
appreciate the influence of rotation on Nu, it is important to understand the dependence of Nu on Ra without rotation.  
The investigation of non-rotating convection has been extensive over almost 40 years with early work focused on classical theories \cite{Malkus54,Howard66,Kraichnan62} that predicted a power law relationship of the form Nu = A Ra$^\beta$ with $\beta = 1/3$. Later measurements, particularly those in helium gas, suggested a value $\beta = 2/7$ with theory and early numerical simulations providing a solid basis for such a law.  A detailed review of these results was presented by Siggia \cite{Siggia94}.   An extension \cite{GL00,GL02} of the competing kinetic and thermal boundary layer theory 
\cite{Shraiman-Siggia90}  that included an expanded analysis of competing boundary and bulk dissipation processes produced a phase diagram with crossover effects between different regions.  Such an approach suggested a form Nu = a Ra$^{\beta_1}$ + b Ra$^{\beta_2}$ with specific predictions for the coefficients derived from fitting a few data sets in each region.  In this latter regard, high precision experimental data for room temperature fluids \cite{Xu00,Nikolaenko03} have been extremely valuable in elucidating differences between a single power law description and one involving two power laws with fixed coefficients.  The measurements presented here are over a modest range of Ra $< 10^9$, and thus cannot distinguish between these two forms.  We compare our results with a broad range of measurements of heat transport of non-rotating convection as a benchmark for considering the effects of rotation on turbulent heat transport.
        
The effects of rotation on convection, especially on heat transport, might be expected to be
substantial given that rotation profoundly changes the nature of boundary layer instability and
modifies the length scales over which motions occur. Whereas thermal plumes are formed in long
sheets and are swept across the cell by mean flow, rotation spins up these plumes into intense
vortical structures. Furthermore, rotation is known  to shorten the linear
length scale dramatically as rotation is increased \cite{Chandra61}. Additional ingredients introduced by
rotation are the Ekman pumping/suction imposed by the differential rotation of the boundary and
the interior flow and the dynamical constraints imposed by the Taylor-Proudman Theorem for
strongly rotating flows. Despite these interesting factors, previous heat transport
measurements have not been well understood for a number of reasons. Rossby, in his
seminal paper on rotating convection \cite{Rossby69}, reported comprehensive heat transport for water and for
mercury as function of Ra and Ta with emphasis on the regions close to onset and of moderate Ra
($< 3 \times 10^6$). His measurements, as well as later measurements in helium \cite{Pfotenhauer87}, quantitatively showed that the convective onset was below the
theoretical prediction of linear stability analysis \cite{Chandra53,Chandra61}. This reduction
in critical Rayleigh number was attributed to a transition to azimuthally-periodic modes
localized near the wall \cite{Buell83,Pfotenhauer87} but neither of
the experiments had flow visualization capabilities. The existence of such wall states was
later  confirmed \cite{Zhong91,Zhong93} using shadowgraph flow visualization
but rather than being stationary  the sidewall states were observed to precess in the rotating
frame counter to the direction of rotation.  This resolved one difficulty with the
data set of Rossby.  

Other experiments \cite{Boubnov86,Boubnov90,Fernando91} and numerical simulations \cite{Raasch91,Jones93,Klinger95} of rotating convection  involve
an open upper fluid surface where one can visualize the development of
convective structures and the interaction of vortices and characterize some of the statistics of
the temperature and velocity fields. These experiments are, however, not amenable to the measurement of 
accurate heat transport. 

The heat transport experiments \cite{Zhong93} that motivated this work used a water in a cylindrical cell with top and bottom rigid boundaries, and measurements were made to higher Ra ($\approx 2 \times 10^7$) than Rossby. The normalized heat transport Nu at constant rotation rate appeared, however, to asymptote to the non-rotating
result at high Ra. This combination of rather small Ra and an apparently non power-law scaling
was confusing.  Numerical simulations \cite{Julien96b} showed, however, that Nu
scaled approximately as the 2/7 power for a fixed convective Rossby number with no-slip top and bottom
boundary conditions. To test this prediction and to further characterize heat transport as a function of rotation, it was necessary to extend the heat transport measurements to higher Ra than in \cite{Zhong93}.   Single point measurements
of temperature made in the same cell as the one used here will be presented elsewhere.
 
This paper is organized as follows. The experimental apparatus and procedures are decribed in
Sec.~\ref{sec:setup}. The heat transport results of non-rotating and rotating turbulence are presented in
Sec.~\ref{sec:nr} and Sec.~\ref{sec:r}, repectively.  Sec.~\ref{sec:con} summarizes the paper.

\section{Experimental procedure}
\label{sec:setup}

\subsection{Rotating apparatus and cell}
\begin{figure}[ht]
\vspace{1cm}  
\resizebox{85mm}{!}{
\includegraphics{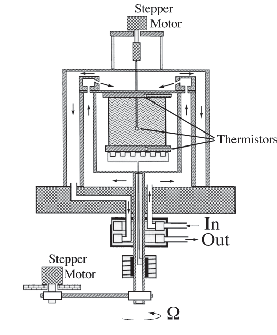}
}
\caption{Schematic of the experimental apparatus}
\label{fig:setup}
\end{figure}

The experimental apparatus, shown schematically in Fig.~\ref{fig:setup},  was an improved
version of the one used previously in studies of rotating convection \cite{Zhong93,Ning93}. The top plate temperature of the cell was controlled by water flow which
was distributed evenly by a set of 12 turrets divided into two groups pointing to 1/3 and 2/3
of the radius, respectively. 
The water flow was temperature-controlled by a refrigerator-circulator and then fed into the
rotating frame through a water slip connection. The flow was further temperature 
regulated by a feedback temperature control unit which maintained the 
top-plate temperature constant with r.m.s. fluctuations of 
less than 1 mK. A film heater attached to the bottom plate 
provided a constant heat current 
to the fluid layer. The power input was obtained by measuring the voltage
across the film heater and the current through it. The latter was 
obtained from  the voltage across a temperature-controlled standard resistor. 

All electrical wires were fed into the rotating frame through an electrical slip
ring and inside a hollowed steel shaft which also served as the drive 
train for rotation. The electrical noise of the slip ring was small enough that there was no
measurable difference in the signals with or without rotation. Rotation
was provided by a micro-stepping motor through a shaft, two gears and
a timing belt and was under computer control. The gear ratio set a lower limit for the frequency of about 0.01 Hz. The maximum frequency surveyed was 0.5 Hz. 

The convection cell was constructed with aluminum top (1.27 cm thick) and bottom (0.64 cm
thick) plates and plexiglass sidewalls (0.32 cm thick). The  aluminum plates were anodized to
prevent corrosion in water. The cell had a height $d=9.40$ cm and horizontal dimensions $L_x =
L_y = 7.30$ cm with an aspect ratio of $\Gamma = L_x/d =0.78$.  The cell's bottom and sides
were insulated by 2.5-cm-thick styrofoam to reduce thermal losses from radiation and conduction
or convection by air.   Four thermistors were embedded in each of the top and bottom plates and
the thermistor centers were located at 0.32 and 0.23 cm from the fluid, respectively,
 and at the mid point between the cell's center and the sides. The four thermistors in each plate gave the average plate
temperature.  The wires from the film heater and from the four thermistors were heat sunk on
the bottom of the cylindrical can which was maintained at the same temperature as the top plate.

\subsection{Heat Transport}

Heat is transported more efficiently by convection where heat can be advected by the fluid
motion than by conduction where heat is transported solely by diffusion. The enhancement of
thermal transport by convection is characterized by the Nusselt number Nu $= K_{eff}/K$
where $K_{eff}$ and $K$ are the effective thermal conductance and molecular thermal conductance
of the fluid layer, respectively. In an experimental realization of convection there are
additional heat transport contributions to the measured thermal conductance. These include
conduction through the cell sidewalls and from conduction, convection, and radiation to the
surrounding environment. Below the onset of convection, the background conductivity can be
measured and effectively subtracted from the total heat transport contribution provided that
the thermal conductivity of  water, available from the literature, is assumed. For turbulent
convection in room temperature experiments this is difficult to accomplish because if  $\Delta
T \approx 10 K$ corresponds to the maximum Ra $\approx 10^9$ then the onset of convection
occurs for $\Delta T \approx 2 \times 10^{-5}K$, far below experimental resolution. This can be
overcome by rotating the cell, thereby suppressing the onset of convection by about four orders
of magnitude, {\it i.e.}, $\delta T \approx 0.2 K$, and allowing the background thermal conductivity
to be measured directly.  We now describe this procedure in detail.

The background thermal conductance was determined to be $K_b = 0.0423 \pm 0.0004$ W/K,
comparable to the water layer's thermal conductance $K = 0.0341$ W/K at a cell mean temperature
of 21.5 $^\circ$C.  Assuming that $K_b$ is independent of the mean temperature of the cell, the
Nusselt number is given by
\begin{equation}
Nu = \frac{\dot{Q}/\Delta T - K_b}{K},
\label{eq:N}
\end{equation}
where $\Delta T = T_b - T_t$ is the temperature difference across the water layer with $T_t$
and $T_b$ being the top-plate and bottom-plate temperatures, $\dot{Q}$ is the total heat input
to the bottom plate, and $K$ is the  thermal conductance of the water layer at the mean
temperature $T_0 = (T_t + T_b)/2$.  The top and bottom temperatures are corrected for
the temperature drop in the aluminum plates although this correction was always
less than 0.3\%.

In our experiment, we measured $K_b$ at a single temperature because the temperature dependence
of the background terms is quite small and does not affect the data presented here. To evaluate the
systematic error in our measurements of $K_b$, we have estimated the different contributions in
that quantity. The major contributors to $K_b$ are the
plexiglass side walls, the insulating foam, the electrical wires, and thermal radiation. The
first three contributions have very weak temperature dependence (less than 1\% change over a
20 K temperature difference)  and small magnitude, estimated to be 0.006,
0.006, and 0.003 W/K, respectively. The rest of the measured background heat transport, about
60\% of $K_b$ ($\approx 0.027$ W/K), comes from a series combination of conduction through the foam, convection in the air surrounding the insulation, and
from thermal radiation from the outer surfaces of the insulation to the surroundings which are
maintained at the top-plate temperature.  The top-plate
temperature was held constant and the bottom-plate temperature was changed by a maximum  of 20
K  corresponding to the maximum heat input. This produces about a 2 K increase in the average
temperature of the radiating surfaces, or about a 2\% increase in radiated heat, which results
in about a 0.02 overestimate in Nu. Therefore, for the measurements with fixed $T_t$, $K_b$ can
be  taken to be constant as the correction to Nu is less than 0.1\%. 

This analysis neglects an important point regarding the heat transported through
the side walls  \cite{ahlers00}.  Rather than supporting a linear temperature profile as in the non-convecting state where the background is measured, the side walls are in contact with a turbulent fluid that is approximately iso-thermal in the bulk of the flow.  This leads to an enhancement of heat transport through the side walls in the turbulent state.  For conditions similar to those presented here (thin plexiglass walls relative to the lateral extent of the system), however, the correction to the total heat transport was shown by numerical modeling to be small, ranging from about 2\% for Nu = 10 to 1\% at Nu = 100.   Our sidewall is thicker by about a factor of two, so in the worst case these values might be twice as large.  Applying
a correction of this order shifts the exponent of a power law fit by at most 1\% (higher) and the constant term by about 7\% (lower).  These estimates contribute to the systematic error in our results but we do not explicitly correct for the side wall
effect in the data presented below.

\subsection{Parameter space}

\begin{figure}[ht]
\vspace{1cm}  \resizebox{85mm}{!}{
\includegraphics{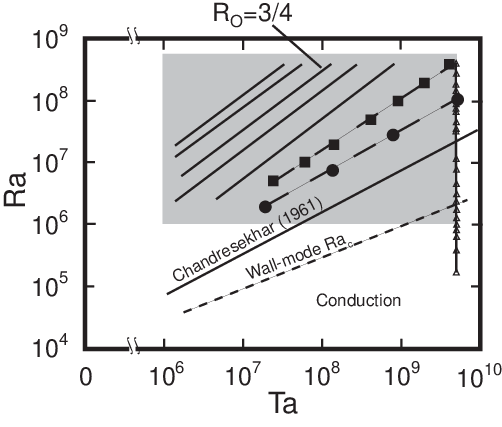}
}
\caption{Parameter space of Ra vs Ta. Most measurements were conducted in the gray
area by fixing Ra, $\Omega_D$, or Ro.  The measurements at $\Omega_D =0.0$ Hz and 3.14
Hz (Ta $\approx 5.0\times 10^9$) started at Ra as low as $5\times 10^5$. Five solid lines in
the gray area represent five different Ro (from right to left, 0.30, 0.52, 0.75, 1.15, and
1.49)  used in the measurements. Data ($\bullet$), from heat transport
measurements at constant $\Omega_D$, set the lower bound of Ra above which Nu could be expected
to exhibit turbulent convection.  Symbols ($\sqbull$), deduced
from Nu data at fixed Ra, represent the loci of maximum Nu at constant Ra. The data ($\bigtriangleup$) at the
highest Ta $\approx 5 \times 10^9$ spanned the largest Ra number range.  The theoretical
prediction of convective onset under  rotation \cite{Chandra61} is also shown in the
figure. }
\label{fig:ra-ta-map}
\end{figure}

The parameter space for rotating convection is defined by Ra and Ta (or $\Omega$) which are
proportional to the physical control variables of $\Delta T$ and $\Omega_D$, respectively. In
Fig.~\ref{fig:ra-ta-map}, the parameter space is shown over a range which encompasses our experimental
measurements. The shaded region denotes the area in which most of our efforts were focused. 
The limitations which determine that area are the range of measurable $\Delta T$ for a given
cell height d; a four-decade variation of Ra was obtained by varying the temperature difference
across the cell from 2 mK to 20 K.  The Taylor number for the shaded area ranged from about $1
\times 10^6$ to $5 \times 10^9$, corresponding to rotation frequencies from $f = 0.01$ Hz to
$0.5$ Hz.  The lower limit was determined by the range of stability of the stepping motor given
a particular gear ratio. By reducing that ratio, lower Taylor numbers could have been
investigated but this proved unnecessary as the interesting range of Ta was spanned with the
chosen gear ratio. 

During the experiments, we  fixed some parameters and studied the dependence of measured
quantities on the others. For instance, we fixed Ra to study the dependence on Ta, and visa
versa. One important parameter, the convective Rossby number Ro which provides a relative
measure of buoyancy relative to rotation, was maintained fixed by varying both $\Delta T$ and
$\Omega_D$ for each data point.  The contours of constant Ro, plotted in
Fig.~\ref{fig:ra-ta-map}, are approximately straight lines in the log-log plot.   
Numerical simulations of turbulent rotating convection with $\sigma =1$ \cite{Julien96a,Julien96b} 
followed the contour of Ro = 3/4 where buoyancy and rotation had roughly comparable
importance. This particular line is noted in the figure. Several other sets of data shown in
the figure are discussed later. Overall,  the parameter ranges in our experiments fall roughly into the region studied in open-top experiments \cite{Boubnov86,Boubnov90,Fernando91}; the parameter space of experiments on rotating convection prior to about 1990 was summarized in \cite{Fernando91}. (the {\it flux} Rayleigh number used in that parameter space  is related to Ra by Ra$_f$ = NuRa so that the highest Ra in our experiments, where Nu $\approx 60$ corresponds to  Ra$_f \approx 3 \times 10^{10}$). 


\section{Heat Transport in Non-Rotating Convection}
\label{sec:nr}
In this section, we present experimental results for non-rotating convection. This  enables us
to compare our results with existing theories and with other non-rotating convection
experiments, of which there are many. It also serves as a reference for our results on rotating
convection. We concentrate here on measurements of heat transport in fluids with Prandtl number $\sigma \approx 6$. 
 
Heat transport is measured by the Nusselt number Nu and scales with Ra as a power-law: Nu $
= A $Ra$^\beta$. Classical arguments \cite{Malkus54,Howard66} suggest $\beta
= 1/3$ whereas other scaling theories \cite{Castaing89,Shraiman-Siggia90}
predict $\beta = 2/7$.  Another approach that generalizes the latter theories \cite{GL00} predicts
a complicated phase diagram as a function of Ra and $\sigma$.  In different regions different power-law scalings dominate with strong crossover effects so that one obtains a form for the heat transport where
Nu scales with Ra as  $aRa^{\beta_1} + bRa^{\beta_2}$ with $\beta_1$ and $\beta_2$ fixed.  Careful heat transport measurements have demonstrated that this latter prediction yields better fits to experimental data \cite{Xu00}. For
the limited range of Ra presented here, either approach yields a good fit to the data.
\begin{figure}[ht]
\vspace{1cm}  \resizebox{85mm}{!}{
\includegraphics{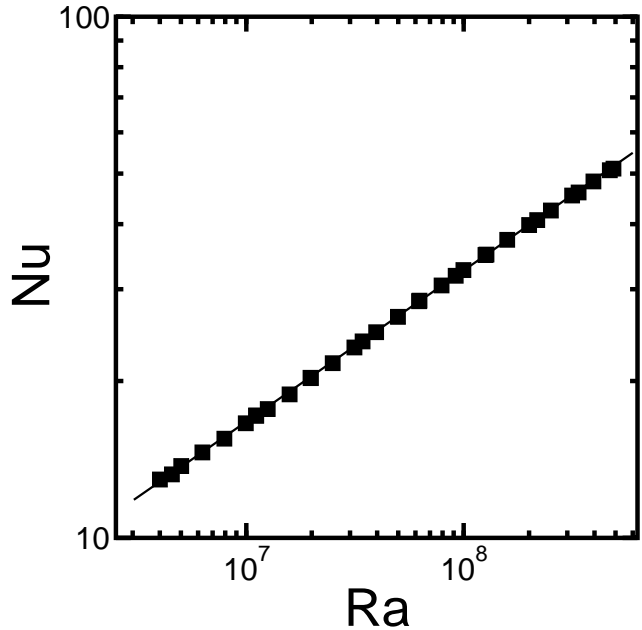}
}
\caption{Nu vs Ra for non-rotating convection. The solid line is the power-law fit 
Nu = 0.158~Ra$^{0.289}$ over the range $4 \times 10^6 < $Ra$ < 5 \times 10^8$. An equivalent fit
is Nu = $0.262\ Ra^{1/5}+ 0.0473\ Ra^{1/3}$.}
\label{fig:nu-omega-0}
\end{figure}

The Nusselt number of non-rotating convection measured with fixed $T_t$ is shown in
Fig.~\ref{fig:nu-omega-0} as a function of Ra for $4\times 10^6 < $ Ra $ < 5\times 10^8$. The
results are reasonably well decribed by a power-law with scaling coefficients 
$A = 0.158 \pm 0.003$ and $\beta =0.289 \pm 0.002$ which were obtained by fitting all data for
$4\times 10^6 < $ Ra $ < 5\times 10^8$ shown in the figure. Fitting the data between $4\times
10^7$ and $5\times 10^8$ yields slightly different values: $A=0.164\pm 0.003$ and $\beta =
0.286\pm 0.002$. These latter values have the virtue of being derived from a range of Ra which
is in the turbulence regime but with the disadvantage of a shorter scaling range.
To account for this systematic uncertainty we take 
$A=0.164\pm 0.006$ and $\beta =  0.286\pm 0.003$ as the best estimates for the  scaling
coefficients. The exponent is very close to  2/7 and
agrees well with earlier work as summarized in Table~\ref{tab:coef}. Although the 2/7 value
has been shown not to describe heat transport data over a much larger range of Ra \cite{Xu00},
we use it here for convenience.  Further, the uncertainties associated with the exponents and 
coefficients are the result of statistical fits and almost certainly underestimate the systematic errors
associated with the different experiments.
\begin{table}
	\begin{center}
		\begin{tabular}{@{}lllllll@{}}
\\ $A$ & $\beta$ & $A_{2/7}$ & $A_{2/7}$ & Ra Range & $\Gamma$ & Reference
\\ & & ($10^7$) &($10^8$) &($\times 10^6$) & &
\\
\\0.131 & 0.300(5) & 0.165 & 0.170 & $0.03 - 2$ & 2.8-10 (C) & Rossby (1969)
\\0.183 & 0.278 & 0.162 & 0.159 & $0.3 - 100$ & 1.5 $\times$ 2.5(R) & Chu \& Goldstein (1973)
\\0.145 & 0.29 & 0.155 & 0.156 & $30 - 4000$ & 3.5-14 (S) & Tanaka \& Miyata (1980)
\\0.137 & 0.275(7) & 0.115 & 0.112 & $2 - 200$ & 0.71, 1.6 (S) & Solomon \& Gollub (1991)
\\0.129 & 0.299(3) & 0.160 &  0.164 & $0.1 - 20$ & 2.0 (C) & Zhong {\it et al.}  (1993)
\\0.19(4) & 0.28(1) & 0.172 &  0.170 & $1 - 400$ & 4 $\times$ 1 (R) &
Chill\'a {\it et al.}  (1993)
\\0.145 & 0.292 & 0.160 &  0.162 & $400 - 7000$ & 1.0 (C) & Cioni {\it et al.} (1996)
\\0.16 & 0.281(2) & 0.147 & 0.145 & $800 - 6000$ & 1.0 (C) & Shen {\it et al.}  (1996)
\\0.164(6) & 0.286(3) & 0.164 & 0.164 & $40 - 500$ & 0.78 (S) & This work
\\0.154 & 0.291 & 0.168 & 0.170 & 3000-60000 & 1.0 (C) & Nikolaenko and Ahlers (2003).
		\end{tabular}
	\end{center}
\caption{Values of heat transport scaling parameters: $A$, $\beta$, $A_{2/7}$ (Ra = 10$^7$),
$A_{2/7}$ (Ra = 10$^8$), Ra-Range, $\Gamma$ (C: cylindrical with $\Gamma = $diameter/height, S:
square with $\Gamma = $width/height, R: rectangular with $\Gamma_{\rm x}\times\Gamma_{\rm y}$),
and reference.  All the experiments listed here used water as the working fluid.  The Prandtl number for different experimental conditions varied slightly  but was in the range $4 < \sigma < 7$.}
\label{tab:coef}
\end{table}

Compared to the scaling exponent $\beta$, the coefficient $A$ is quite different from one experiment to another. $A$
is sensitive to the exponent and a precise determination of $A$ requires a larger range of Ra than has been
available in any of the experiments using water. Fixing the exponent at 2/7 and computing a
value for $A_{2/7}$ (equivalent to Nu/Ra$^{2/7}$) at different Ra gives a better comparison
between data sets, see Table~\ref{tab:coef}. For the water experiments and for $10^7 < Ra <
10^9$, the coefficients agree quite well except for the experiments of Solomon and Gollub
\cite{Solomon91} where a liquid mercury bottom surface may account for the discrepancy. All of the data
reported earlier and listed in the Table do not directly measure the background heat transport
contribution that we are able to account for using rotation. This background measurement is important
in eliminating systematic error for smaller Nu.  An average over all the data sets
for convection in water yields $A = 0.161 \pm 0.007$ and $\beta = 0.287 \pm 0.008$ with no
statistically-significant dependence of Nu on $\Gamma$. In summary, our data for non-rotating convection
agree well with earlier results despite the significant variation in aspect ratios between
experiments. 

We also took heat transport data at fixed mean-cell temperature $T_0$ to estimate the Prandtl number
dependence of the heat transport \cite{Liu97} (not reported here).  We can use that data to 
correct the data at fixed $T_t$. In the fixed $T_t$ measurements reported in the remainder of this paper, we had a
variation in the range   21.5$^\circ$C $< T_0 <$ 31.4$^\circ$C (and resultant $6.7>\sigma>5.2$)
corresponding to changes from the lowest to highest heat input. Such non-constant $\sigma$ or
$T_0$ results in an uncertainty in Nu of the order of 0.8\%. We interpolate the Nu data to a constant mean temperature or a
constant $\sigma$. We choose $T_0 = 26.0^\circ$C (where $\sigma=5.93$) as the reference
temperature which was about the average of the mean temperatures in the experiments. The
interpolated value is given by Nu=Nu$_m (\sigma/5.93)^{0.0292}$ where Nu$_m$ is the measured
Nusselt number using Eq.~\ref{eq:N}. The difference between the interpolated and measured
values of Nu is less than 0.8\%, and does not change the scaling coefficients within their
specified error bars.  Nevertheless, the interpolated values are reported in Fig.\ \ref{fig:nu-omega-0}.

One important feature of turbulent convection is the large scale circulation that is driven by an accumulation
of thermal plumes that congregate near the lateral boundaries \cite{Siggia94,AhlersRMP}. The general circulation in our cell was visualized using glass
encapsulated thermochromic liquid crystals (TLC). A white light sheet about 1mm in width was
used to illuminate the cell from the side and a black background was provided for good
contrast.  The mean-flow direction  was typically across the cell diagonal. Flow reversals were
observed as was a shifting of the main diagonal circulation in a clockwise or counterclockwise
direction as viewed from above. This meant that sometimes the flow along the off-diagonal
direction reversed directions frequently.  This cross diagonal flow has been observed before in
convection cells with square cross section \cite{Solomon91}.
In addition to the large diagonal flows there were often small recirculating
cells in the corners and along the bottom-side boundary. 
Viewed from above the thermal plumes near the bottom boundary layer
were arranged into coherent sheets which were swept up by the mean flow. This was also seen in
a number of convection experiments in water where the flow is easy to visualize \cite{Tanaka80,Zocchi90,Solomon91}. 
\section{Heat Transport in Rotating Convection}
\label{sec:r}

Heat transport measurements in the presence of rotation are complicated by the changing
influences of buoyancy, proportional to Ra, and rotation, proportional to $\Omega_D$. The
simplest thing to do experimentaly is to fix $\Omega_D$ and vary $\Delta T$. Because the mean
temperature changes with $\Delta T$ (fixed $T_t$), the dimensionless $\Omega$ changes owing to
the temperature-dependent viscosity of water. This can be as large as 25\% over the Ra range
that we studied. As noted above $\sigma$ also changes somewhat but that influence is small. 
Even constant $\Omega$ is not the appropriate variable to hold constant if one wants
to evaluate the behavior of Nu as a function of Ra for a constant rotational forcing. From previous heat transport measurements \cite{Zhong93}, it appeared that Nu at fixed $\Omega_D$ was enhanced by rotation for intermediate Ra
but seemed to asymptote to the non-rotating value of Nu at higher Ra. A convenient 
measure of rotational forcing \cite{Julien96a,Julien96b} is defined by the ``convective'' Rossby number, Ro$ \equiv \sqrt{{\rm Ra}/(\sigma {\rm Ta})}$, which is the
ratio of  a rotational period to a buoyancy time. So Ro $ \approx 1$ should mark the border
between strongly rotating convection with Ro $ << 1$ and weakly rotating convection with Ro $ >>
1$. In those numerical simulations of rotating convection at
fixed Ro = 0.75, Nu scaled approximately as Ra$^{2/7}$.  Here we present heat transport 
measurements to higher Ra than previously \cite{Zhong93} and consider Nu as a function of Ra at fixed Ro. In addition, we
compare our results to earlier ones by Rossby \cite{Rossby69} who compiled constant Nu contours as a
function of Ra and Ta. To get a good understanding of the whole system, it is useful to
consider different slices of the parameter space. We present them in the order of contours of
Nu, fixed $\Omega_D$, fixed Ra, and finally fixed Ro. 

\begin{figure}[ht]
\vspace{1cm}
\resizebox{85mm}{!}{
\includegraphics{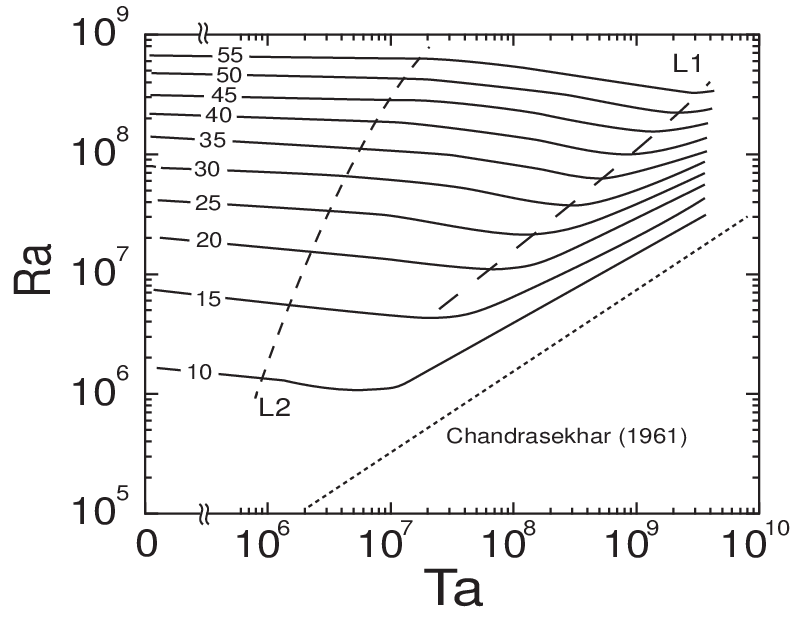}
}
\caption{Lines of constant Nu as a function of Ra and Ta. The dotted line is for the
bulk convective onset from Chandrasekhar and Ra$_c \propto {\rm Ta}^{2/3}$. Line $L1$
connects the loci of maximum Nu at constant Ra and agrees well with the loci of the minimum Ra
at constant Nu. Line $L2$ delimits the parameter space into the right section where we made
measurements under rotation and the left section which are interpolated between zero rotation
and the data with the lowest Ta.}
\label{fig:iso-nu}
\end{figure}

Heat transport measurements in rotating convection can be summarized by a contour plot of Nu
presented in Fig.~\ref{fig:iso-nu}. 
The points on each constant-Nu line were obtained by interpolating Nu data  measured under
different controlled conditions (namely, fixed
$\Omega_D$, Ra, and Ro). The individual data points are within 1\% of the smooth curves for
large Nu and within 3\% for the smallest Nu. Figure~\ref{fig:iso-nu} complements the Nu  contour
plot in Fig.~11 of Rossby \cite{Rossby69} where the highest Nu was 12, and the combination of the two
gives a rather complete description of Nu in the parameter space of Ra$_c \leq Ra < 10^9$ and $0
\leq {\rm Ta} < 10^{10}$. As shown in Fig.~11 of \cite{Rossby69} for the lower Ra and Ta range,
there is a minimum Ra for constant Nu or alternately there is a maximum Nu at constant Ra. At
fixed Ta, however, Nu is a monotonically increasing function of Ra. In the following, results
are presented which elucidate the origin of the maximum Nu at fixed Ra (or the minimum Ra at
fixed Nu) and investigate the variation of Nu as a function of Ra, Ta, and Ro. Before 
proceeding with these details, however, we can already see the overall trend of heat transport at fixed
Ra. For low rotation, Nu is rather insensitive to changes in Ta (note the discontinuity in the
horizontal axis in Fig.\ \ref{fig:iso-nu}). In the intermediate range between lines {\it L2} and
{\it L1}, Nu increases with increasing Ta, as rotation enhances heat transport. For high enough
Ta, however, rotation suppresses convection and Nu decreases as the onset of bulk convection is
approached at Ra$_c$(Ta). 

\subsection{Constant $\Omega_D$}

\begin{figure}[ht]
\vspace{1cm}
\resizebox{85mm}{!}{
\includegraphics{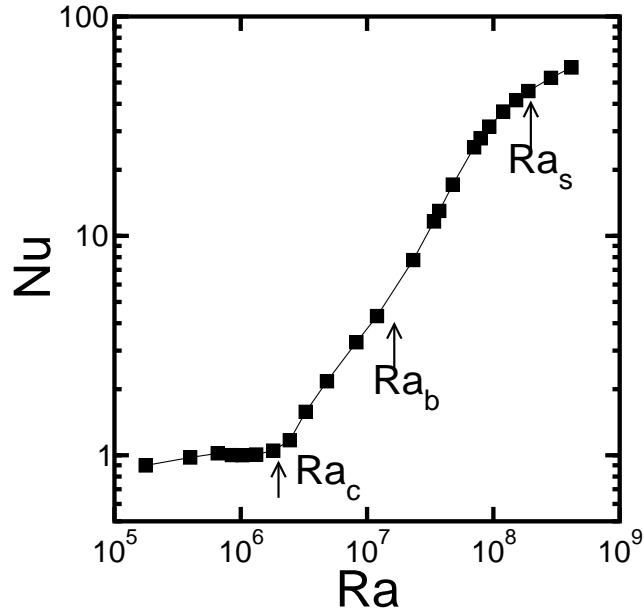}
}
\caption{Nu vs Ra for $\Omega_D = 3.14$ ($\Omega \approx 3.3\times 10^4$). The arrows indicate
the onset of the side-wall traveling state at Ra$_c$ and the bulk state Ra$_b$. The approximate
onset of turbulent convection is indicated as Ra$_s$. The line is a guide to the eye.}
\label{fig:nu-omega-500}
\end{figure}

Shown in Fig.~\ref{fig:nu-omega-500} is Nu versus Ra at $\Omega_D = 3.14$ Hz where we obtained
the background conductance $K_b$ as described in Sec.~\ref{sec:setup}. 
The first few points have larger uncertainty because of the small temperature difference
$\Delta T$ across the cell and the long thermal diffusion time in our cell (about 16 h).  The
transition to  convection  from a conduction state occurred at $\Delta T_c \approx 150$ mK and
Ra$_c \approx 2 \times 10^6$, much lower than the theoretical value of $1.6\times 10^7$ for a
laterally-infinite system at this rotation rate \cite{Chandra61}. This lower-than-expected
transition was observed in early heat-transport experiments \cite{Rossby69,Pfotenhauer87,Zhong93} and was later visually identified as a transition to a
side-wall traveling-wave state \cite{Zhong91,Zhong93}. Extrapolating the results in
\cite{Zhong93} to $\Omega = 2.88 \times 10^4$, one obtains the onset to
the traveling state at about $3\times 10^6$, not far from our value of $2\times 10^6$. There
is also an inflection point in Nu at Ra$_b \approx 1.6 \times 10^7$, coinciding with the
theoretical prediction \cite{Chandra61} for the transition to bulk convection \cite{Zhong93,Ning93}. Note that as the mean temperature of the
cell increased during the measurements (constant $T_t$), $\Omega$ and Ta increased from $2.88
\times 10^4$ and $3.31 \times 10^9$ at small Ra to $3.58 \times 10^4$ and $5.12 \times 10^9$ at
the highest Ra, respectively. 

\begin{figure}[ht]
\vspace{1cm}  \resizebox{85mm}{!}{
\includegraphics{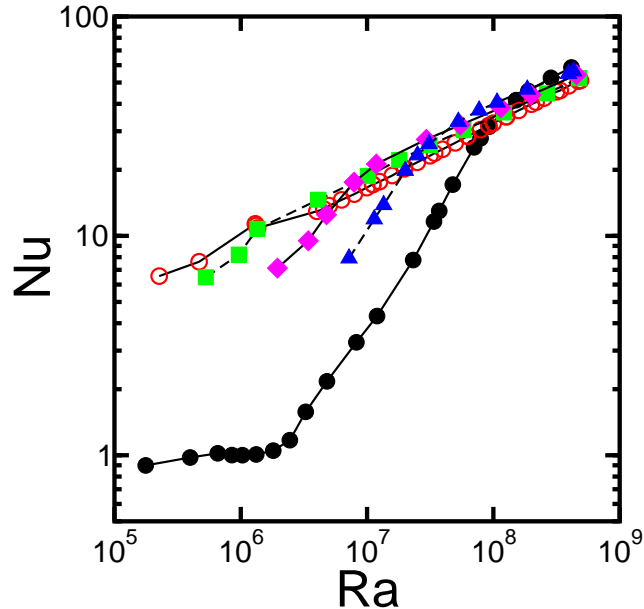}
}
\caption{(Color online) Nu vs Ra at constant $\Omega_D$: 0.00($\circ$), 0.188($\sqbull$), 0.502($\diamond$),
1.26($\bigtriangleup$), and 3.14($\bullet$). 
 The lines are guides to the eye.}
\label{fig:nu-fix-omega}
\end{figure}

We also measured Nu at $\Omega_D = 0.00$, 0.188, 0.502, and 1.26 Hz. These measurements,
shown in Fig.~\ref{fig:nu-fix-omega},  served as a rough characterization of the system. 
The first observation is that higher rotation suppresses convection relative to non-rotating
convection from onset up to a value of Ra that depends on rotation. Above that Ra, Nu is
higher than its corresponding value without rotation. Thus, the notion that rotation is a
damping influence on convection, as suggested by the Taylor-Proudman theorem, is only valid
near onset and the opposite is true for turbulent convection.  The values of Ra$_s$ for this
crossover are plotted in  Fig.~\ref{fig:ra-ta-map} as solid circles. It is the crossover that gives rise to the
maxima in the contours of Nu. The second thing to notice is that although
Nu(R,$\Omega$)/Nu(Ra,0) $> 1$  at intermediate Ra, it appears to asymptote to 1 at higher Ra.
This suggests that buoyancy wins out over rotation at high Ra and fixed $\Omega$. 

Revisiting the presence of mean flow, a feature of non-rotating convection over a large range of Ra,
we consider mean flow for rotating convection.  On the one hand, flow visualization in the rotating frame for a cylindrical cell with $5 \times 10^7 < R < 5 \times 10^8$ \cite{Vorobieff02}  used both thermochromic liquid crystal (TLC) and particle image velocimetry to determine the flow  structure near the upper boundary layer \cite{Vorobieff02}). The
sheet-like plumes evolved under rotation into vortices and for small enough Ro, {\it i.e.}, for
rotation dominated flow,  there was no indication of a large scale circulation extending over
the size of the container. It seems that the shear on the boundary layer is of a very different form than for
non-rotating convection as strong vortical motions dominate the flow just outside the boundary
layer.  Results for higher Ra in the range $10^9$ to $3 \times 10^{11}$ \cite{Hart02} indicated a precessing mean
flow provided Ro $> 0.5$.  Similarly, recent results showed a breakdown of large-scale circulation for Ro $< 1.2$ \cite{Kunnen08}.

\subsection{Constant Ra}
\begin{figure}[ht]
\vspace{1cm}  \resizebox{85mm}{!}{
\includegraphics{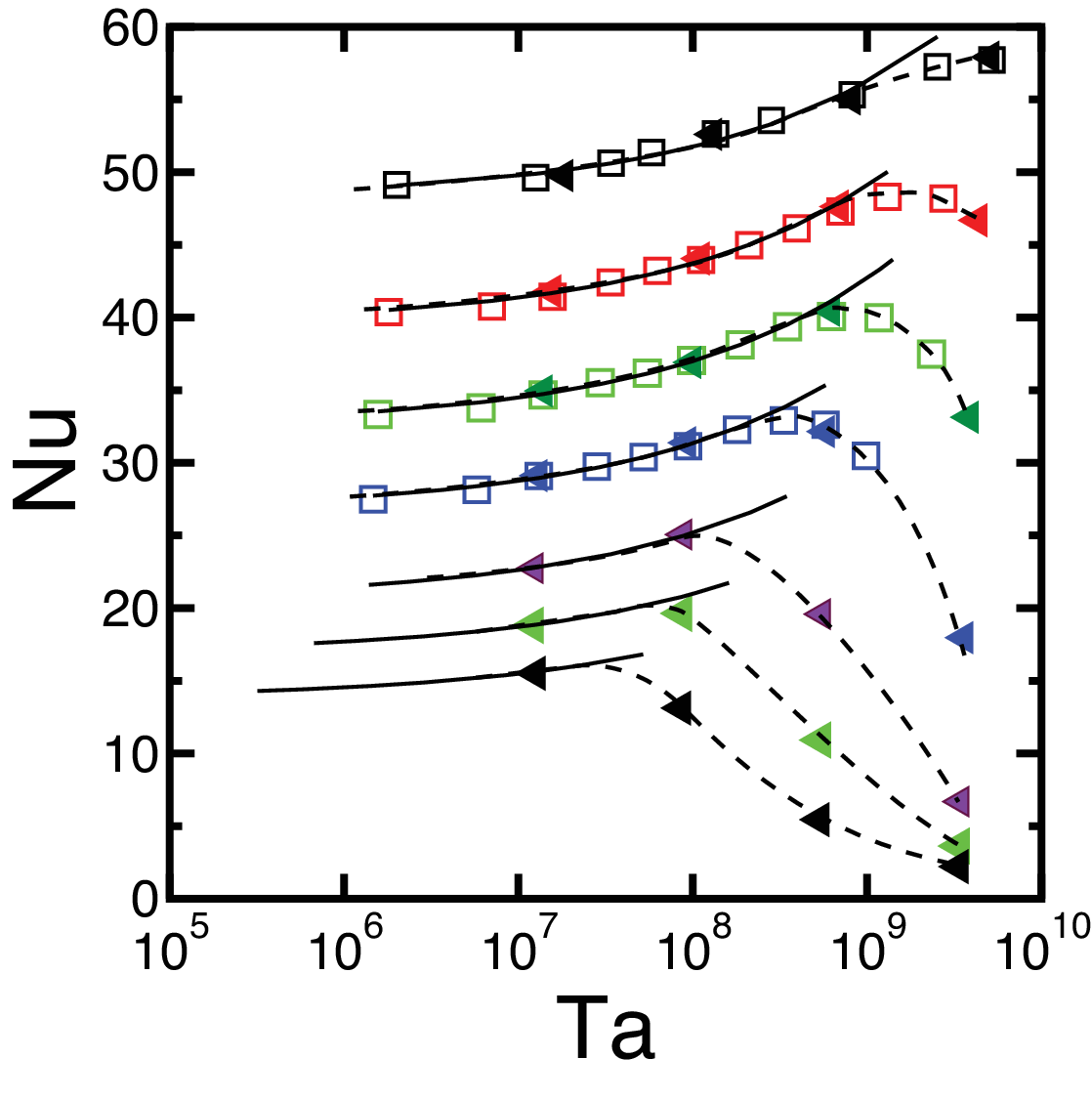}
}
\caption{(Color Online) Nu vs Ta at (from top to bottom) Ra$=4.0 \times 10^8$, $2.0 \times 10^8$, $1.0 \times
10^8$, $5.0 \times 10^7$, $2.0 \times 10^7$, $1.0 \times 10^7$, and $5.0 \times 10^6$. $\square$
-- measured experimentally with Ra held fixed and $\lhd$ -- interpolated from data measured at constant $\Omega_D$
(Fig.\ \protect{\ref{fig:nu-fix-omega}}). Solid lines are least-square fits to the data; dashed lines are guides to
the eye.}
\label{fig:Nu-Ta-fixRa}
\end{figure}

Another way to look at the influence of rotation on convection is to fix Ra and vary $\Omega$.
We measured Nu as a function of Ta at Rayleigh numbers of Ra $= 5\times 10^7, 1\times 10^8, 
2\times 10^8, 4\times 10^8$, with results as shown in 
Fig.~\ref{fig:Nu-Ta-fixRa}. 
For each Ra, the data were corrected so that $\sigma$  was constant. We have also included data
for Ra $= 2\times 10^7, 1\times 10^7,  5\times 10^6$ which were obtained by interpolating data
measured at constant Ro and $\Omega_D$.  For all the data sets, Nu increased with rotation
before decreasing at higher Ta with a maximum Nu that varied with Ra. The loci Ta$_{m}$ of
maximum Nu are plotted in Fig.~\ref{fig:ra-ta-map} as solid squares  and in
Fig.~\ref{fig:iso-nu} (line $L1$). Note  in the latter figure that this line  approximately
connects the loci of minimum Ra for each Nu contour. Line $L1$ can be approximated  by Ra $
\approx 2.2 {\rm Ta}^{0.85}$, differing from the relation of Ra $ \approx 206 {\rm Ta}^{0.63}$ at
lower Ra and Ta reported by Rossby (1969). The very different exponents indicate a
continuously steepening curve and  suggests that there is no clear asymptotic (in Ta) power-law
scaling for the Nu maxima over the Ta range studied so far.  

\begin{figure}[ht]
\vspace{1cm}  \resizebox{85mm}{!}{
\includegraphics{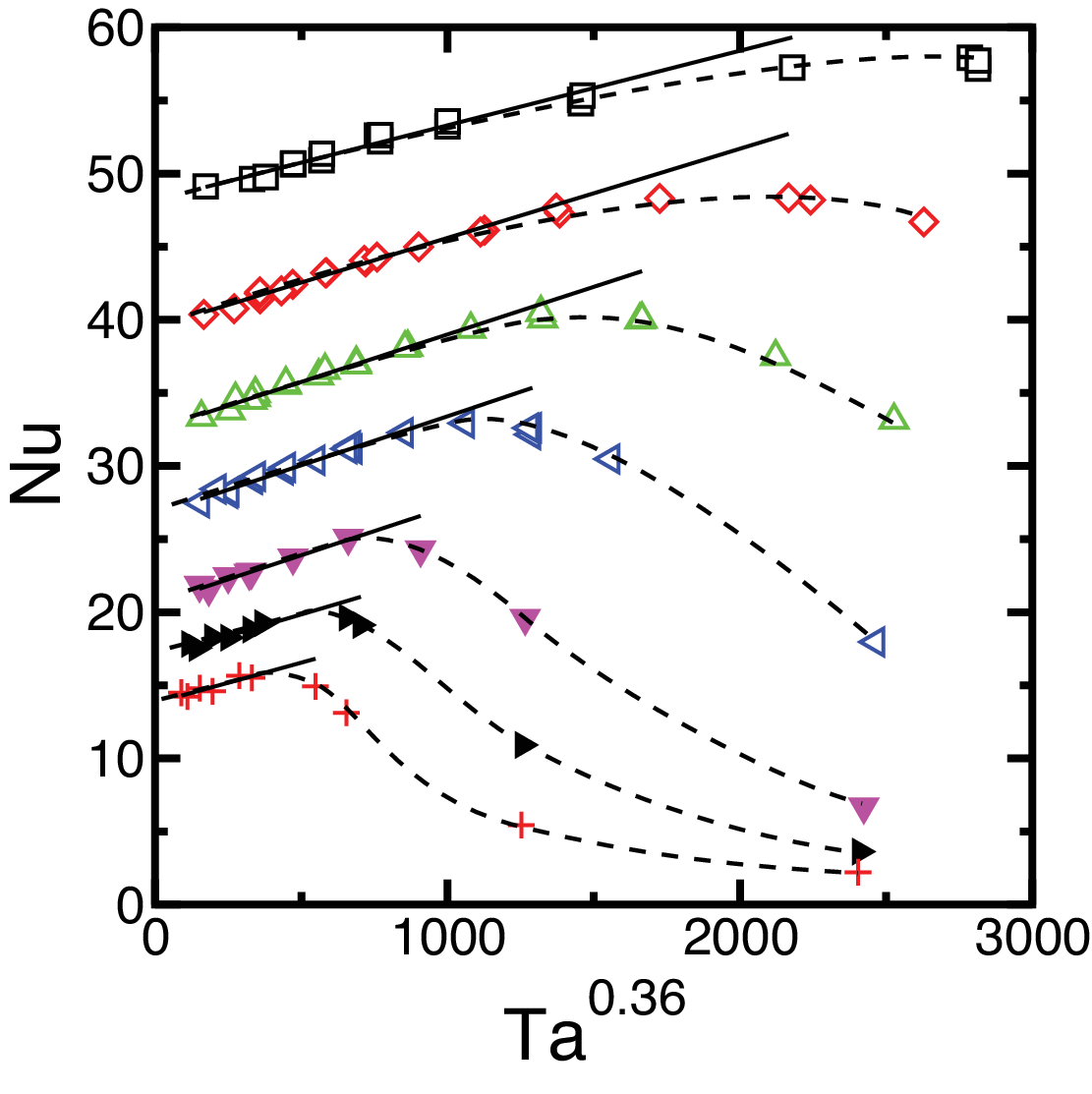}
}
\caption{(Color online) Nu vs Ta$^{0.36}$ so that least-squared fits (solid lines) of Nu=a Ta$^{0.355}$ yield straight lines. 
From top to bottom, Ra$=4.0 \times 10^8$, $2.0 \times 10^8$, $1.0 \times 10^8$, $5.0 \times 10^7$, $2.0 \times 10^7$, $1.0
\times 10^7$, and $5.0 \times 10^6$. The dashed lines are guides to the eye and show deviation
from linear fit.}
\label{fig:Nu-Ta-fixRa-fit}
\end{figure}

An interesting conjecture regarding the enhancement of heat transport by rotation is that
rotation creates thermal vortices which increase Nu through Ekman pumping in the boundary layer
\cite{Zhong93,Julien96b}. Thus, the enhancement in Nu might be
proportional to the number of such vortices. Because we have not visualized the flow for the
rotating system, the number of vortices as a function of Ta is not known directly \cite{Sakai97,Vorobieff02}.
 Instead we consider the linear prediction for the number of structures at the convective onset. 
 The linear wavenumber $k_c$
increases with Ta and asymptotically scales like Ta$^{1/6}$ \cite{Chandra61} which implies
that the number of cellular structures should scale like $k_c^2\ \sim\ {\rm Ta}^{1/3}$.  This
scaling for vortex number was observed even significantly above onset in experiments with an
open top surface \cite{Boubnov86} which suggests it is a reasonable assumption here.
Instead of the 1/3 scaling of the asymptotic theory, however, we will compare with an empirical fit to the
linear data over our range of Ta which gives $k_c(Ta)/k_c(0) \approx 0.090 Ta^{0.355}$.  In
Fig.~\ref{fig:Nu-Ta-fixRa-fit}, we plot Nu in Fig.~\ref{fig:Nu-Ta-fixRa} as a function of
Ta$^{0.36}$. 
For Ta $ < $ Ta$_m$, there is a linear region for each Ra that shrinks as Ra decreases. In the
linear region, we have 
\begin{equation}
{\rm Nu} = {\rm Nu}_0 + \gamma {\rm Ta}^{0.355},
\end{equation}
where Nu$_0$ and $\gamma$ are fitting parameters and values for different Ra are listed in
Table~\ref{tab:tab1}. 
\begin{table}
	\begin{center}
		\begin{tabular}{@{}ccccccc@{}}
\\Ra & Nu$_m$ & Ta$_m$ & Nu$(\Omega_D=0)$ & Nu$_0$ & $\gamma$
\\($\times 10^6$) & & ($\times 10^7$) & & &
\\400  & 58.0 & 420   & 48.38 & 48.2(2)   & 0.0051(5)
\\200  & 49.0 & 200   & 39.87 & 39.5(1)   & 0.0061(4)
\\100  & 40.5 & 90     & 32.62 & 32.5(1)   & 0.0065(2)
\\50    & 33.0 & 42     & 26.60 & 26.7(1)   & 0.0067(3)
\\20    & 25.5 & 14     & 20.31 & 20.6(2)   & 0.0066(4)
\\10    & 20.5 & 6       & 16.62 & 16.9(2)   & 0.0059(9)
\\5      & 16.0 & 2.5    & 13.65 & 13.8(3)   & 0.0055(12)
		\end{tabular}
	\end{center}
\caption{Fitting parameters Nu$_0$ and $\gamma$.  Nu$_m$ and Ta$_{m}$ are the maximum Nusselt
number and its location at constant Ra. Nu$(\Omega_D=0)$ is the non-rotating value obtained by
cubic least-square interpolation of  $Log$(Nu) vs $Log$(Ra) for non-rotating convection.}
\label{tab:tab1}
\end{table}
Solid lines in Figs.~\ref{fig:Nu-Ta-fixRa} and \ref{fig:Nu-Ta-fixRa-fit} are calculated from
these fitting parameters. The deviation from the solid lines (linear behavior) at higher Ta is
a result of rotation suppressing convection in the weakly nonlinear regime near onset. Also
listed in Table~\ref{tab:tab1} is Nu$(\Omega_D =0)$, the Nusselt number of non-rotating
convection. The fitting parameter Nu$_0$ for all Ra is nearly identical to Nu$(\Omega_D =0)$ 
within  fitting and experimental uncertainties. This indicates that the fitting is consistent
with the data in the range 0 $<$ Ta $<$ Ta$_{m}$. Thus, the enhancement of Nu by rotation is
given by $\Delta$Nu = $\gamma$Ta$^{0.355}$ with $\gamma$ about 0.006. This result supports the
conjecture that the enhancement is proportional to the average number of thermal vortices.

Using velocity field measurements \cite{Vorobieff02}, the number of vortices could be evaluated
more quantitatively.  At a fixed Ra = $3.2 \times 10^8$, Ta was varied from 0 up to about $10^{10}$.
The variation of vortex (cyclonic) density with $\Omega$ was slightly sub-linear, over the range
$10^7 < {\rm Ta}  < 10^9$.  Since $\Omega \sim Ta^{1/2}$, this result suggests that the number of vortices
depends on Ta with a power a bit less than 1/2. This is roughly consistent with the estimate based
on linear stability arguments.  Unfortunately, the vortex density data are too sparse to provide better
estimates in the Ta range of interest.

\subsection {Constant {\rm Ro} -- power-law scaling}

Many experiments in thermal convection without rotation showed that Nu scales more closely
as the 2/7 power of Ra in the regime $10^7 < Ra < 10^9$ than with the classical 1/3 power
law. As discussed earlier, a generalized theory in terms of a phase diagram in Ra and $\sigma$ and more precise experimental measurements suggest a form with the sum of two power laws with exponents of 1/5, 1/4, 1/3, or 1/2 depending on the region in phase space.  Rotation complicates the issue of scaling since the relative influence
of rotation changes with changing Ra at fixed Ta.  Numerical simulations \cite{Julien96b} showed that the convective Rossby number is a good measure of the relative
importance of buoyancy with respect to rotation: for Ro = 0.75 and $\sigma =
1$,  Nu $ \sim$ Ra$^{2/7}$ which indicates that the details of rotation are relatively
unimportant in the determination of the scaling behavior. We have tested this prediction and
over the range $0.1 < {\rm Ro}  < 1.5$ find approximate 2/7 power-law scaling.  We use this single
power law description for convenience -  a fit of the form Nu = a Ra$^{1/5}$ + b Ra$^{1/3}$ yields equivalent
fits.  For the non-rotating case, the coefficients are a = 0.26 and b=0.047

\begin{figure}[ht]
\vspace{1cm}  
\resizebox{85mm}{!}{
\includegraphics{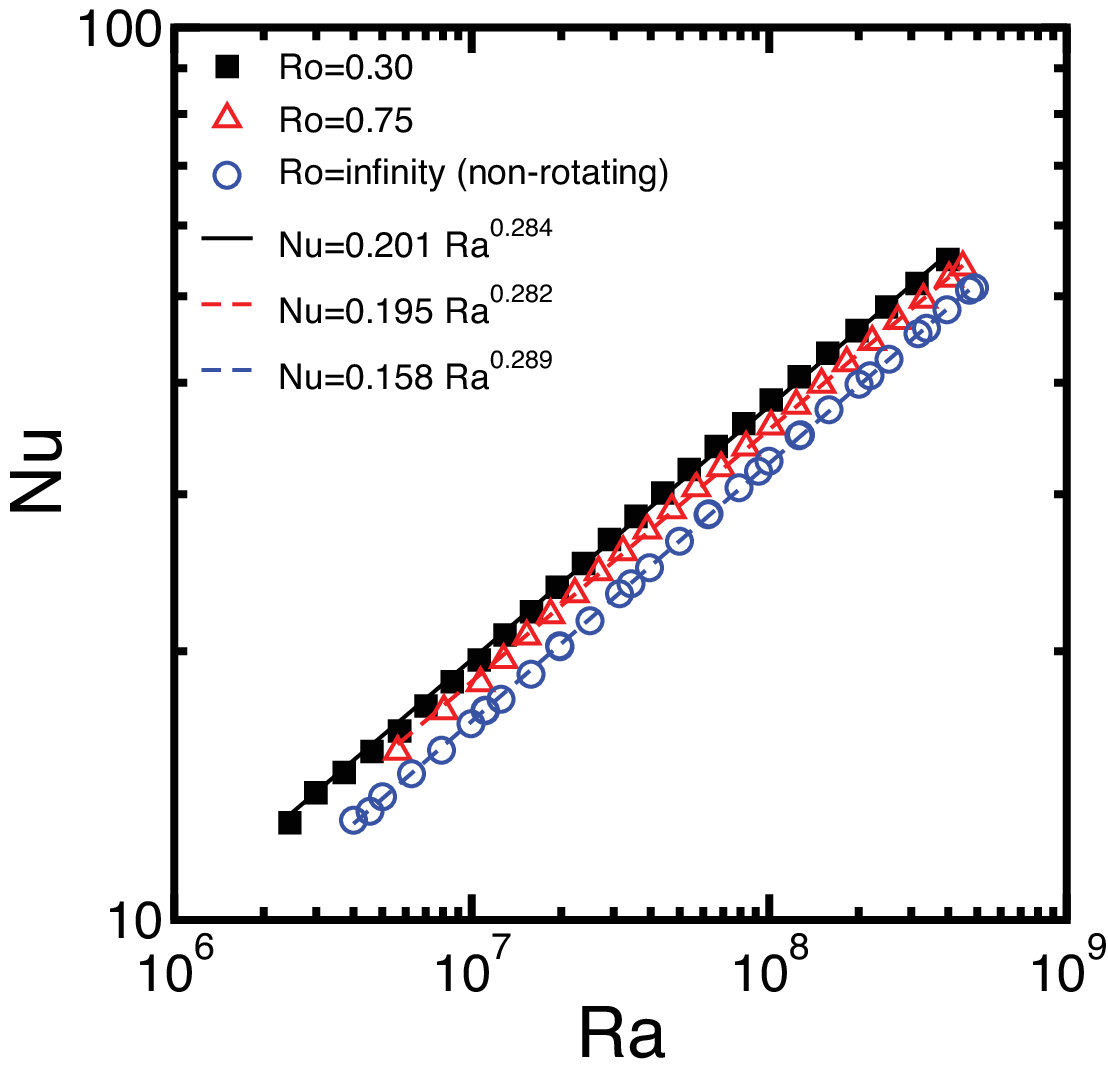}
}
\caption{(Color online) Nu vs Ra at constant Ro: 0.30($\sqbull$), 0.75($\bigtriangleup$), and infinity($\circ$). The
dashed lines are power-law fits with amplitudes and exponents  listed in the legend. }
\label{fig:nu-fix-Ro}
\end{figure}

\begin{figure}[ht]
\resizebox{85mm}{!}{
\includegraphics{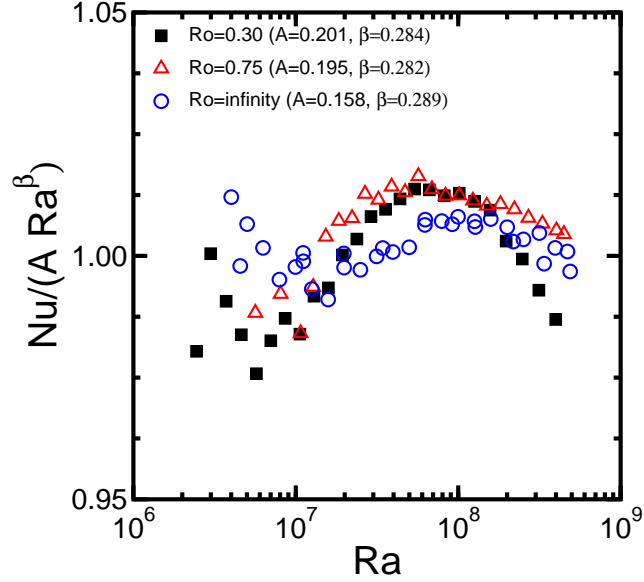}
}
\caption{(Color online) Nu/Ra$^\beta$ vs Ra at constant Ro: 0.30($\sqbull$), 0.75($\bigtriangleup$), and
infinity($\circ$, non-rotating).}
\label{fig:nu-fix-Ro-scaled}
\end{figure}

\begin{figure}[ht]
 \resizebox{85mm}{!}{
\includegraphics{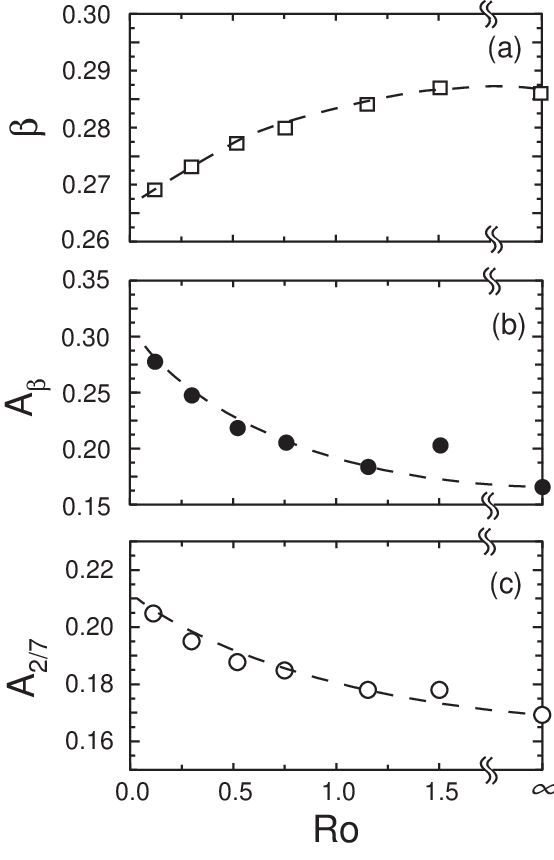}
}
\caption{Rotation-dependence of the fitting parameters: (a) $\beta$ and  (b) $A_{\beta}$ in
Nu $= A_{\beta} Ra^{\beta}$, and (c) coefficient $A_{2/7}$ in Nu $= A_{2/7} Ra^{2/7}$. Dashed lines
are guides to the eye. }
\label{fig:A-a-Ro}
\end{figure}

We have measured Nu at several constant Rossby numbers, i.e., Ro = 0.12, 0.30, 0.52, 0.75, 1.15,
1.49, and $\infty$ (zero rotation). For clarity, only part of the data for Ro = 0.30, 0.75, and
infinity are plotted on a log-log scale in Fig.~\ref{fig:nu-fix-Ro} with the coefficients of power law fits listed in the plot.  
The Nusselt number agrees well with a 2/7 power law and is different from the 1/3 power law,
especially at higher Rayleigh number ($> 4 \times 10^7$). In Fig.~\ref{fig:nu-fix-Ro-scaled}, we
plot Nu/Ra$^{2/7}$ versus Ra to gauge  how well the 2/7 power law  describes the data. 
Over the Ra range where the 2/7 power law is satisfied, a constant value of $A_{2/7} =
Nu/Ra^{2/7}$ is expected. The description is reasonably good for the non-rotating case, but
becomes a little worse as Ro decreases or rotation increases. The coefficient $A_{2/7}$ is
plotted in Fig.~\ref{fig:A-a-Ro}(c). The trend of increasing $A_{2/7}$ with decreasing Ro demonstrates the enhancement of
heat transport by rotation.
 
We also fit the data with Nu $= A_{\beta} Ra^{\beta}$ to obtain the coefficient and the exponent
as functions of Ro. Least-squares fitting was performed in the range $4\times 10^{7} < Ra <
5\times 10^{8}$ to avoid possible deviation from power-law scaling at lower Ra. The results are
plotted in Fig.~\ref{fig:A-a-Ro}. It should be pointed out that for non-rotating convection the
coefficient $A = 0.164$ and exponent $\beta = 0.286$ are slightly different from the values
obtained by fitting the data for $4\times 10^{6} < {\rm Ra} < 5\times 10^{8}$. For finite Ro, the
exponent depends on Ro: it decreases almost monotonically from 0.287 at Ro = 1.5 to 0.269 at
Ro = 0.12.  As shown in Figs.~\ref{fig:nu-fix-Ro},~\ref{fig:nu-fix-Ro-scaled},
and~\ref{fig:A-a-Ro}, the coefficient $A$ increases as rotation increases, and its value is
very sensitive to the fitting value of exponent as can be seen by comparing
Fig.~\ref{fig:A-a-Ro}(b) and (c). 

Determining unambiguously the scaling behavior of the heat transport requires many orders of
magnitude in Ra. Thus, an absolute comparison of scaling exponents in our experiment is
uncertain.   Nonetheless, our experiments yielded some interesting results, especially when
compared to numerical simulations \cite{Julien96a,Julien96b}.  First, at fixed Ro,
Nu depends on Ra with a power law close to 2/7, in agreement with numerical simulation \cite{Julien96b}
for Ro = 0.75 and $\sigma =1$. If anything, rotation seems to reduce the
scaling exponent slightly. This could be the result of different scaling ranges as a function
of Ro, because fixed Ro does not exactly maintain a balance between buoyancy and rotation, or
because rotation modifies the scaling exponent directly. An extended range in Ra would be
necessary to resolve this quantitatively, perhaps in a gas system.  

From the perspective of turbulent convection theory \cite{Castaing89,Shraiman-Siggia90,GL00}, 
the insensitivity of the scaling to rotation is rather interesting because 
rotation affects many properties of the turbulence, such as the change from thermal plumes to
vortices and the existence of a turbulent Eckman boundary layer and associated Eckman
pumping.   In non-rotating convection, the relationship between the thermal boundary layer
thickness $\delta_T$ and the viscous sublayer thickness $\delta_{\nu}$ determines the power-law
scaling in the sheared boundary layer theory \cite{Shraiman-Siggia90} where Nu $ \sim$ Ra$^{2/7}$ applies when
$\delta_T < \delta_{\nu}$. Rotation introduces another vertical length scale, the Ekman layer
thickness $\delta_E$, which could in principle play a role similar to  $\delta_{\nu}$ in
non-rotating convection. 

\begin{figure}[ht]
\vspace{1cm}  \resizebox{85mm}{!}{
\includegraphics{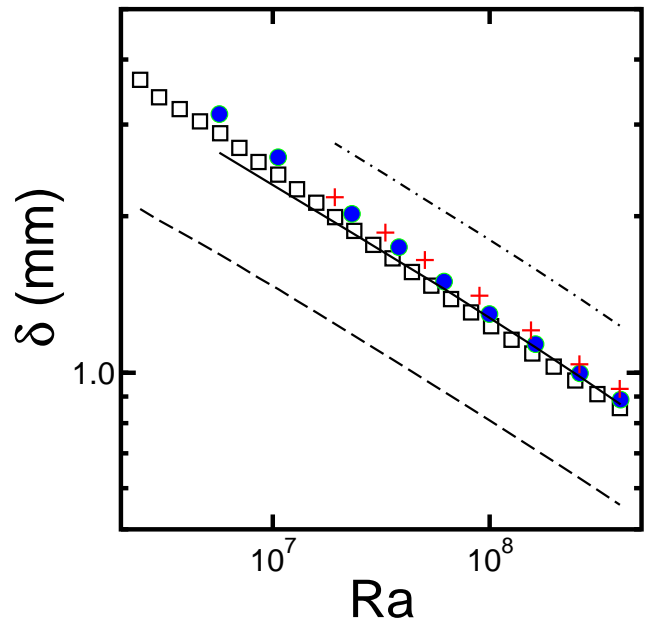}
}
\caption{(Color online) Calculated thermal boundary layer thickness ($\delta_T$, symbols) and Ekman layer
thickness ($\delta_E$, lines) at Ro=0.30 ($\square$, dashed line), 0.75 ($\bullet$, solid line),
and 1.49 ($+$, dotted-dashed line).}
\label{fig:bl-scaled}
\end{figure}

In Fig.~\ref{fig:bl-scaled}, we plot $\delta_T$ and  $\delta_E$ as functions of Ra at Ro=0.30,
0.75, 1.49 where $\delta_T$ and  $\delta_E$ are defined as 
\begin{eqnarray}
\delta_T & = & {1 \over 2} {d \over Nu} \\ 
\delta_E & = & \left (\frac{\nu}{2\Omega_D}\right )^{1/2} = \frac{d}{{\rm Ta}^{1/4}}
\end{eqnarray}
Numerical simulations \cite{Julien96a} showed that  the transition to turbulent scaling occurs
at Ra $\approx 4\times 10^7$ for Ro=0.75 where $\delta_T \approx \delta_E$. 
In our experiment we also have $\delta_T \approx \delta_E$ at Ro = 0.75 despite the different Prandtl number
used in the experiment.  Thus, the condition $\delta_T \approx \delta_E$ for a
turbulence transition is not consistent with our experiments. It is also interesting to notice
that at high Ra $ >> 10^8$ and constant Ro, $\delta_T \propto R^{-2/7}$ decreases faster than
$\delta_E\ \sim $ Ra$^{-1/4}$, therefore
$\delta_T$ and $\delta_E$ cross at high Ra for Ro $<$ 0.75 and do not cross at Ra for Ro $>$
0.75. The interplay of these two length scales is as yet not understood nor is it clear that
$\delta_E$ defined from non-convecting problems with differential rotation is the proper
variable to use here. It is perhaps interesting to note that along lines of constant Ro, the
ratio $\delta_T/\delta_E$ is almost constant.  A direct measurement of the turbulent Ekman
layer in the presence of a thermal boundary layer would be very useful to augment the arguments
based on numerical simulations \cite{Julien96a} for a coexisting thermal boundary layer with a linear Ekman layer. 
Subsequent flow visualization \cite{Vorobieff02} using particle image velocimetry showed an diverse set of interesting
behaviors of velocity and vorticity fluctuations but could not yield a definitive conclusion regarding the complex
interplay of thermal and kinetic boundary layers involved in determining heat transport for turbulent convection.


\section{Conclusions}
\label{sec:con}

We have presented experimental studies of turbulent thermal convection in water confined in a
cell with a square cross section with and without rotation. In non-rotating convection, the
Nusselt number was found to scale roughly as Ra$^{2/7}$ at $4\times 10^6 < Ra < 5\times 10^8$. 
Heat transport measurement in rotating convection confirmed the findings by other researchers \cite{Rossby69,Zhong93}
that rotation enhances thermal transport over a certain
range of Ra and Ta range. Evidence showed that such enhancement may be  attributed to the
increased effective horizontal area caused by the presence of vortices under rotation. At fixed
Rossby number, Nu was found to scale approximately as Ra$^{2/7}$, as does Nu of non-rotating convection and
predicted from numerical simulation \cite{Julien96b}. Characterization by a combination of power laws \cite{GL00,GL02}
was equally good at fitting the data for both rotating and non-rotating convection.

Analysis using Ekman layers instead of kinetic boundary layers as input into a scaling theory did not
provide additional insight into the heat transport data, and it remains unclear how rotation and its associated
modification of boundary layer structure affects heat transport.  Given the extensive study of non-rotating
convection in recent years with great advances in characterizing boundary layers, heat transport and large scale circulation \cite{AhlersRMP}, there seems to be an emerging opportunity to apply similar rigor to the geophysically important case of rotating thermal convection.  We hope that our work may stimulate more experimental and theoretical studies on rotating  thermal turbulence. A rigorous test of the power-law scaling of Nu under constant Ro and the validity
of Ro as the 'good' parameter requires a much larger Ra and Ta range than what was available in
our experiment. A gas convection may be needed to achieve such experimental conditions. The
interplay amongst various length scales, such as horizontal vortex wavelength, Ekman layer,
thermal boundary layer, viscous sublayer, was not well understood in our experiment and certainly
calls for further work on this aspect of turbulent convection. 

\begin{acknowledgements}
We would like to thank Joe Werne, Keith Julien, Peter Vorobieff, Phil Marcus for helpful
discussions. This work was supported by the U.S. Department of Energy.
\end{acknowledgements}

\noindent $\dag$ Present Address: Jet Propulsion
Laboratory, California Institute of Technology, 4800 Oak Grove Drive, MS 79-24, Pasadena, CA 91109



\end{document}